\documentclass[12pt]{article}
\input epsf.sty

\begin{document}

\def\CA{{\cal A}}
\def\CB{{\cal B}}
\def\CC{{\cal C}}
\def\CD{{\cal D}}
\def\CE{{\cal E}}
\def\CF{{\cal F}}
\def\CG{{\cal G}}
\def\CH{{\cal H}}
\def\CI{{\cal I}}
\def\CJ{{\cal J}}
\def\CK{{\cal K}}
\def\CL{{\cal L}}
\def\CM{{\cal M}}
\def\CN{{\cal N}}
\def\CO{{\cal O}}
\def\CP{{\cal P}}
\def\CQ{{\cal Q}}
\def\CR{{\cal R}}
\def\CS{{\cal S}}
\def\CT{{\cal T}}
\def\CU{{\cal U}}
\def\CV{{\cal V}}
\def\CW{{\cal W}}
\def\CX{{\cal X}}
\def\CY{{\cal Y}}
\def\CZ{{\cal Z}}

\newcommand{\todo}[1]{{\em \small {#1}}\marginpar{$\Longleftarrow$}}
\newcommand{\labell}[1]{\label{#1}\qquad_{#1}} 
\newcommand{\bbibitem}[1]{\bibitem{#1}\marginpar{#1}}

\newcommand{\llabel}[1]{\label{#1}\marginpar{#1}}

\newcommand{\sphere}[0]{{\rm S}^3}
\newcommand{\su}[0]{{\rm SU(2)}}
\newcommand{\so}[0]{{\rm SO(4)}}
\newcommand{\bK}[0]{{\bf K}}
\newcommand{\bL}[0]{{\bf L}}
\newcommand{\bR}[0]{{\bf R}}
\newcommand{\tK}[0]{\tilde{K}}
\newcommand{\tL}[0]{\bar{L}}
\newcommand{\tR}[0]{\tilde{R}}

\newcommand{\btzm}[0]{BTZ$_{\rm M}$}
\newcommand{\ads}[1]{{\rm AdS}_{#1}}
\newcommand{\ds}[1]{{\rm dS}_{#1}}
\newcommand{\eds}[1]{{\rm EdS}_{#1}}
\newcommand{\sph}[1]{{\rm S}^{#1}}
\newcommand{\gn}[0]{G_N}
\newcommand{\SL}[0]{{\rm SL}(2,R)}
\newcommand{\cosm}[0]{R}
\newcommand{\hdim}[0]{\bar{h}}
\newcommand{\bw}[0]{\bar{w}}
\newcommand{\bz}[0]{\bar{z}}
\newcommand{\be}{\begin{equation}}
\newcommand{\ee}{\end{equation}}
\newcommand{\bea}{\begin{eqnarray}}
\newcommand{\eea}{\end{eqnarray}}
\newcommand{\pat}{\partial}
\newcommand{\lp}{\lambda_+}
\newcommand{\bx}{ {\bf x}}
\newcommand{\bk}{{\bf k}}
\newcommand{\bb}{{\bf b}}
\newcommand{\BB}{{\bf B}}
\newcommand{\tp}{\tilde{\phi}}
\hyphenation{Min-kow-ski}

\def\apr{\alpha'}
\def\str{{str}}
\def\lstr{\ell_\str}
\def\gstr{g_\str}
\def\Mstr{M_\str}
\def\lpl{\ell_{pl}}
\def\Mpl{M_{pl}}
\def\varep{\varepsilon}
\def\del{\nabla}
\def\grad{\nabla}
\def\tr{\hbox{tr}}
\def\perp{\bot}
\def\half{\frac{1}{2}}
\def\p{\partial}
\def\perp{\bot}
\def\eps{\epsilon}
\newcommand{\Tr}{\mathop{\rm Tr}}


\renewcommand{\thepage}{\arabic{page}}
\setcounter{page}{1}

\def\NN{{\cal N}}
\def\nfour{{\cal N}=4}
\def\ntwo{{\cal N}=2}
\def\none{{\cal N}=1}
\def\nonestar{{\cal N}=1$^*$}
\def\tr{{\rm tr\ }}
\def\RR{{\cal R}}
\def\PP{{\cal P}}
\def\ZZ{{\cal Z}}

\newcommand{\bel}[1]{\be\label{#1}}
\def\al{\alpha}
\def\bt{\beta}
\def\mn{{\mu\nu}}
\newcommand{\rep}[1]{{\bf #1}}
\newcommand{\vev}[1]{\langle#1\rangle}
\def\bra{\langle}
\def\ket{\rangle}
\def\eref{(?FIX?)}

\rightline{hep-th/0111163}
\rightline{UPR-967-T}
\vskip 1cm
\centerline{\Large {\bf Giant Gravitons and a Correspondence Principle}}
\vskip 0.5cm
\renewcommand{\thefootnote}{\fnsymbol{footnote}}
\centerline{{\bf Vijay
Balasubramanian\footnote{vijay@endive.hep.upenn.edu} and
Asad Naqvi\footnote{naqvi@rutabaga.hep.upenn.edu}
}}
\centerline{\it David Rittenhouse Laboratories, University of
Pennsylvania}
\centerline{\it Philadelphia, PA 19104, U.S.A.}

\setcounter{footnote}{0}
\renewcommand{\thefootnote}{\arabic{footnote}}

\begin{abstract}
We propose a correspondence between the physics of certain small
charge black holes in $\ads{k} \times \sph{l}$ and large charge
black holes in $\ads{l} \times \sph{k}$.  The curvature
singularities of these solutions arise, following Myers and
Tafjord, from a condensate of giant gravitons.  When the number of
condensed giants $N_g$ is much greater than the number of
background branes $N$, we propose that the system has an
equivalent description in terms of $N$ giant gravitons condensed
in a background created by $N_g$ branes. Our primary evidence is
an exact correspondence between gravitational entropy formulae of
small and large charge solutions in different dimensions.
\end{abstract}

\section{Introduction}
 String theory enjoys a number of duality symmetries that
 relate the physics of apparently different theories in
 different dimensions. A particularly striking illustration
 of such a symmetry is the AdS/CFT duality \cite{Maldacena},
which relates gravity on AdS space to a CFT on the AdS boundary.
In this note we propose a correspondence between the physics of
large charge black holes in $\ads{k} \times \sph{l}$ and small
charge black holes in $\ads{l} \times \sph{k}$.

The black holes in question are electrically charged under $U(1)$
gauge fields arising from Kaluza-Klein reduction of 10d or 11d
supergravity on a sphere.  Myers and Tafjord~\cite{myers} showed that
the curvature singularities of such charged solutions of $\ads{5}
\times \sph{5}$ gravity can be understood as condensates of giant
gravitons~\cite{giant} on $\sph{5}$.
In Sections~2 and~3 we extend the work of Myers and
Tafjord~\cite{myers} to charged black holes in the four and seven
dimensional gauged supergravities.  Lifting the solutions to
eleven dimensions~\cite{manyauthors} as asymptotically
$\ads{k}\times\sph{l}$ spaces reveals that their singularities
arise from condensation of giant gravitons on $\sph{l}$.  In four
and seven dimensions there are four and two possible U(1) charges
that the black holes can carry, while the five dimensional
solutions in~\cite{myers} can carry three charges.  Each of the
different charges arises from a different species of giant
graviton.  In all cases, the single charge BPS solution has a
horizon that coincides with the curvature singularity and adding
some energy creates a solution with finite gravitational entropy.

As is well known, $\ads{4}\times\sph{7}$, $\ads{5}\times\sph{5}$
and $\ads{7}\times\sph{4}$ arise in string theory as the near
horizon geometries of stacks of M2, D3 or M5-branes.  Typically,
we take the near-horizon limit of a flat brane and find the
Poincar\'e patch of $\ads{k}$, and a duality follows with the low
energy world-volume CFT of the corresponding planar $k-2$ brane.
The charged black holes studied here are in global AdS space, so
we do not have such a near-horizon construction.\footnote{However,
see~\cite{BdBRK} for global $\ads{3}$ arising from the
near-horizon limit of the spinning D1-D5 string.  It would be
interesting if a similar construction could be carried out in
higher dimensions from the near horizon limit of spinning branes.}
Nevertheless, these spaces are dual to a $k-2$ brane CFT on a
sphere. Hence we will speak of $N$ spherical ``background'' branes
creating the spacetime.  As described above, the single charge
black hole solutions of these theories can be interpreted as
condensates of giant gravitons which are themselves spherical
branes moving on the sphere factor of $\ads{k} \times \sph{l}$.
The giants of $\ads{4}$, $\ads{5}$ and $\ads{7}$ are spherical M5,
D3 and M2 branes respectively. Near any one of these branes, the
geometry should locally be $\ads{l}\times \sph{k}$. When the
number of giants is very large, the background
spacetime should be dominated by the presence of the giants rather
than the presence of background branes. 
This leads to an intriguing hypothesis: {\it When the number of
condensed giants $N_g$ is much greater than the number of
background branes $N$, the system has an
equivalent description in terms of $N$ giant gravitons condensed
in a background created by $N_g$ branes.}

In Section~4 we find evidence for such a correspondence by
examining the thermodynamics of near-BPS, single charge black
holes of $\ads{k} \times \sph{l}$.   We measure entropies and
temperatures in terms of the number of condensed giants, and find
an exact match between the large charge black holes in
$\ads{k}\times\sph{l}$ and the small charge black holes in
$\ads{l} \times \sph{k}$ when the number of giants is exchanged
with the number of background branes while holding energies and
the AdS scale fixed.  Implications of our findings and directions
forward are discussed in Section~5.

\section{Four dimensions}

The Kaluza-Klein compactification of M-theory on $S^{7}$ can be
truncated consistently to $SO(8)$ gauged ${\cal N}=8$ supergravity in
four dimensions.  We are interested in black holes charged under the
maximal Abelian subgroup $U(1)^{4}$.  There is a truncation of the
full $\CN = 8, \, SO(8)$ theory to a $U(1)^{4}, \,  \CN=2$ theory, for which
the bosonic fields are the metric $g_{\mu \nu}$, four $U(1)$ gauge
fields $A_i$, three scalars and three pseudo-scalars (the pseudo-scalars
will be set to zero in this note).  This theory admits four-charge AdS
black hole solutions, given by~\cite{4dholes}:
\begin{eqnarray}
ds_4^2&=&-(H_1H_2H_3H_4)^{-1/2} f \, dt^2 + (H_1H_2H_3H_4)^{1/2}
(f^{-1}dr^2+r^2 d\Omega_{2}^2), \nonumber \\ X_i& = &
H_i^{-1}(H_1H_2H_3H_4)^{1/4}, \,\,\,\,\,\,\,\,
A^i=\frac{\tilde{q}_i}{r+q_i} \, dt, ~~~~~(i = 1\cdots 4)
\label{4dbh}
\end{eqnarray}
where
\[
f=1-\frac{\mu}{r}+\frac{r^2}{L_4^2}(H_1H_2H_3H_4), ~~~~~~
H_i=1+\frac{q_i}{r}.
\]
The four $X_i$ satisfy the relation $X_1X_2X_3X_4=1$,
parameterizing three physical scalars, while $\mu$ is a SUSY
breaking parameter.  The BPS solution occurs when $\mu=0$.  The
physical $U(1)$ charges $\tilde{q_{i}}$ are given in terms of
$q_{i}$ as
\begin{equation}
\tilde{q}_i=\sqrt{q_i(\mu+q_i)} \,. \label{physch4}
\end{equation}
The mass of this black hole is:
\begin{equation}
M_4 = \frac{1}{4G_4}(2\mu +\sum_{i=1}^4 q_i) \equiv
\frac{1}{4G_4}\sum_{i=1}^4 q_i + 2 \, \delta M_4 \,,\label{MADM4}
\end{equation}
where $2\delta M_4$ measures the deviation from the BPS limit.  Black
holes carrying a charge under one of the four $U(1)$'s have a
curvature singularity surrounded by horizon, which shrinks to zero
area as $\mu\rightarrow 0$.  The multi-charge solutions are
qualitatively different.  They have a critical value,
$\mu=\mu_{crit}$, below which the horizon disappears giving a nakedly
singular space.  For $\mu > \mu_{crit}$, there is a regular black hole
horizon with finite entropy and a non-zero temperature.  As
$\mu\rightarrow\mu_{crit}$ from above, the horizon approaches a finite
limiting area.  This critical black hole has zero temperature, but
finite entropy~\cite{therm}.

\paragraph{Lifting to 11D supergravity:}
The general Kaluza-Klein ansatz that lifts any solution of 4D $\ntwo,
U(1)^4$ gauged SUGRA to a solution of 11D supergravity was given in
\cite{manyauthors}.  The charged black hole solution (\ref{4dbh})
lifts to
\begin{eqnarray}
ds_{11}^2& = &
\Delta^{2/3}\Bigl((H_1H_2H_3H_4)^{-1}f\, dt^2+(f^{-1}dr^2+r^2
d\Omega_2^2)\Bigr)\\ && \,\,\,\,\,\,\,
+4\Delta^{-1/3}\sum_i H_i\Bigl(L_4^2 \,d\mu_i^2+\mu_i^2 \, (L_4 \, d\phi_i+
\frac{\tilde{q}_i}{r+q_i}dt)^2\Bigr) \,, \nonumber
\end{eqnarray}
where $\sum \mu_{i}^{2}=1$, $\Delta=(H_1H_2H_3H_4)\sum_{i=1}^4
(\mu_i^2 / H_i)$, and $d\Omega_{2}^{2} =\sin^2 \alpha_1
d\alpha_{1}d\alpha_{2}$ is the volume element on a unit
two-sphere. The solution is asymptotically $\ads{4} \times
\sph{7}$ where the AdS and sphere length scales are $L_4$ and
$2L_4$ respectively. The sphere is parameterized by $\phi_{i}$ and
$\mu_{i}$ as
\begin{equation}
d\Omega_7^2=\sum_i (d\mu_i^2+\mu_i^2 d\phi_i^2)\,,
\end{equation}
and the four $\mu_{i}$ are expressed in terms of three angles
as
\[
   \mu_1=\cos\theta_1, ~
\mu_2=\sin\theta_1\cos\theta_2, ~
\mu_3=\sin\theta_1\sin\theta_2\cos\theta_3, ~ {\rm and} ~
\mu_4=\sin\theta_1\sin\theta_2\sin\theta_3. \] The lift of the
black hole also has a four form field strength $F^{(4)}=dB^{(3)}$
with
\begin{equation}
B^{(3)}=-\frac{r^3}{L_4}\Delta \, dt \wedge d\Omega_{2}^{2}
-L_4 \sum_{i=1}^{4}q_i \mu_i^2(L_4 \, d\phi_i-dt)\wedge
d\Omega_{2}^{2} \, .
\end{equation}

\paragraph{Interpretation as condensed giants: }
The lifted 11 D black hole solutions have curvature singularities
localized in $\ads{4}$ and distributed all over $\sph{7}$.  These
singularities are even naked in the multi-charge cases for small
$\mu$.  In the analogous $\ads{5} \times \sph{5}$ solutions, Myers and
Tafjord~\cite{myers} argued that the singularity could be understood
as a condensate of giant gravitons, by showing that the 5-form flux
near the singularity in their solution was consistent with presence of
a distribution of spherical D3-branes on $\sph{5}$.  Then, using the
charge-mass relation of BPS giant gravitons they showed that the BPS
solutions have an ADM mass that is also consistent with a source that
is a condensate of giant gravitons~\cite{myers}.

In our case, the relevant brane source will be a distribution of giant
gravitons on $\sph{7}$.  These are spherical M5 branes occupying
an $\sph{5}$ of the $\sph{7}$ and carrying angular momentum along one
direction of the sphere~\cite{giant}.  We will start with the BPS case
($\mu =0$) and a single non-zero charge ($q_1$) and then state the
results for the general case.

Being spherical branes, our giants correspond to M5 brane dipoles.
They locally excite the seven-form field strength which can be detected
by integrating the dual four-form over a surface enclosing a part of the
M5-sphere.    A closed surface transverse to the M5-brane spans the
angular coordinates of $\ads{4}$ and the directions on $\sph{7}$ that
are transverse to the M5-brane. The relevant four-form component is
therefore
\begin{equation}
F^{(4)}_{\theta_1 \phi_1 \alpha_1 \alpha_2}=4{q}_1L_4^2
\sin\theta_1 \cos\theta_1 \sin\alpha_1 \, .
\end{equation}
Integrating this form over the angular coordinates transverse to
the giant gravitons ($\alpha_{1,2}$ in AdS and $\phi_{1}$ and
$\theta_{1}$ on the sphere) at any fixed $r$ and $t$ gives a net
flux which is proportional to the number of enclosed giants.
Following~\cite{myers,giant}, giant gravitons exciting these
four-form components are moving in the $\phi$ direction of
$\sph{7}$ and localized along the $\theta$ direction.  We can
express this number in terms of 11d Planck length $l_{p}$ and $N$
which counts the number of units of background 4-form flux that
are present independently of the giant gravitons, or equivalently
the number of M2-branes whose near-horizon limit yields $\ads{4}
\times \sph{7}$ with AdS length scale $L_4
=l_p(\frac{1}{2}\pi^2N)^{1/6}$.  Then, with our conventions,
\begin{equation}
16\pi G_{11} T_5 n_1 = \int d\theta_1 d \phi_1 d\alpha_1 d\alpha_2 \,
F_{\theta_1 \phi_1 \alpha_1 \alpha_2}\,,
\end{equation}
where $G_{11} =16\pi^7l_p^9$ and $T_5=\frac{2\pi}{(2\pi l_p)^6}$ is
the M5 brane tension.  By dropping the integration over $\theta_1$, we
obtain the distribution of giant gravitons in $\theta_1$:
\begin{equation}
\frac{dn_1}{d\theta_1}=\frac{N^{1/2}}{8\sqrt{2}\pi^2L_4^3} \int
F^{(4)}_{\theta_1 \phi_1 \alpha_1 \alpha_2} d\phi_1 d^2 \alpha
=2N^{1/2}\frac{q_1}{\sqrt{2}L_4} \sin 2\theta_1 \,. \label{4ddist}
\end{equation}
Integrating over $\theta_1$ gives the total number of
giants,
\begin{equation}
n_1=\int_0^{\pi/2} d\theta_1
\frac{dn_1}{d\theta_1}=\sqrt{2N}\frac{q_1}{L_4} \, .
\label{4dtot1}
\end{equation}
Treating a single giant graviton as a probe in a background
$\ads{4} \times \sph{7}$ geometry, one finds that a giant located
at $\theta_1$ has a radius $L_4\sin \theta_1$ and carries an
angular momentum along the $\phi_1$ direction equal to $N \sin^4
\theta_1$~\cite{giant}. Using this relation, we expect that the
total angular momentum of the condensate of giants (\ref{4ddist})
is
\begin{equation}
P_{\phi_1}=\int_0^{\pi/2}d\theta_1 \, \frac{dn_1}{d\theta_1} N
\sin^4\theta_1=\frac{2N^{3/2}}{3\sqrt{2}}\frac{q_1}{L_4}\,.
\label{4dmom}
\end{equation}
Likewise,  using the energy-momentum relation of BPS giant gravitons,
we expect that a condensate with distribution (\ref{4ddist}) has a
total energy
\begin{equation}
E=\frac{P_{\phi_1}}{2L}=\frac{N^{3/2}}{3\sqrt{2}}\frac{q_1}{L_4^2}\,.
\label{4den}
\end{equation}
This agrees beautifully with the ADM mass (\ref{MADM4})
\begin{equation}
M_4=\frac{1}{4G_4}q_1=\frac{N^{3/2}}{3\sqrt{2}}\frac{q_1}{L_4^2} \, ,
\end{equation}
where we have used
$G_4= G_{11}/{\rm Vol}(S^7)=\frac{3G_{11}}{128\pi^4L_4^7}$.

We can extend this analysis to the multi-charge black hole
solutions with arbitrary $q_i$.  The generic BPS solution is
nakedly singular. The singularity arises from condensation of sets
of giant gravitons, separately moving along the different
$\phi_i$.  It is convenient to use the radius of the giant
graviton moving along $\phi_i$ as a coordinate $\rho_i=2L_4
\sqrt{1-\mu_i^2}$.  Then, an analysis like the one given above
shows that the distribution of each set of giant gravitons is
\begin{equation}
\frac{dn_i}{d\rho_i}= 2\sqrt{2N} \frac{q_i}{L_4^3}\rho_i\,.
\end{equation}
The giant graviton of radius $\rho_i$ carries an angular momentum
$N \rho_i^4/L_4^4$. The total angular momentum carried by each set
of giant gravitons is
\begin{equation}
P_i = \frac{2N^{3/2}}{3\sqrt{2}}\frac{q_i}{L_4}\,.
\end{equation}
Likewise, the total energy of the giant gravitons $E=\sum P_i/2L_4$
is in agreement with the ADM mass of the multi-charge black
hole.

\paragraph{Beyond BPS: } So far we have considered the BPS
solutions with $\mu=0$.  As a result, we were able to compute the energy
of a condensate of probe giant gravitons following the distribution
(\ref{4ddist}) and this exactly matched the gravitational mass of the
fully back-reacted spacetime solution.  Repeating the analysis in the
non-supersymmetric case when $\mu > 0$, we  find that the net
giant number density, the net number of giants and their net momentum
are simply given by replacing $q_{i}$ in (\ref{4ddist}),
(\ref{4dtot1}) and (\ref{4dmom}) by the physical charge
$\tilde{q}_{i}$ (\ref{physch4}) of the non-extremal solution.
However, replacing $q_{i}$ by $\tilde{q}_{i}$ in the giant energy
formula (\ref{4den}) does not reproduce the spacetime mass
(\ref{MADM4}) because the non-extremality can produce additional
fluctuations of the giant gravitons as well as giant-anti-giant pairs
at any fixed net momentum (\ref{4dmom}).

\section{Seven dimensions}
In an analysis parallel to the one above, we can consider the $\CN
= 4, \, SO(5)$ gauged supergravity in 7 dimensions, arising from
dimensional reduction and consistent truncation of 11d SUGRA on
$\sph{4}$.  This theory can be further truncated to $\CN=2, \,
U(1)^2$ gauged SUGRA, where the only bosonic fields retained are
the metric, two gauge potentials and two
scalars~\cite{manyauthors}.\footnote{This is not in general a
consistent truncation but is so for solutions of the form
considered here~\cite{manyauthors}.}  The charged black hole
solutions in this theory are
\begin{eqnarray}
ds_7^2&=&-(H_1H_2)^{-4/5} f \, dt^2 + (H_1H_2)^{1/5}
(f^{-1}dr^2+r^2 d\Omega_{5}^2) \nonumber \, ,\\ X_i& = &
H_i^{-1}(H_1H_2)^{2/5} \, , \,\,\,\,\,\,\,\,
A^i=\frac{\tilde{q}_i}{r^4+q_i}\, dt \, , \label{7dbh}
\end{eqnarray}
and
\begin{equation}
f=1-\frac{\mu}{r^4}+\frac{r^2}{L_7^2}(H_1H_2), \,\,\,\,\,\,\,
H_i=1+\frac{q_i}{r^4}, \, ~~~~ i=1,2.
\end{equation}
Here $\mu$ is a SUSY-breaking parameter and the $\tilde{q}_i$ are
defined as in (\ref{physch4}).    The mass of this solution is
\begin{equation}
     M_7 = {\pi^{2} \over 4 G_{7}} ( {5 \mu \over 4} + q_{1} +
     q_{2} )
=
 {\pi^{2} \over 4 G_{7}} (  q_{1} +
     q_{2} ) + 5 \delta M_7
\, , \label{7dmass}
\end{equation}
where $\delta M_7$ measures the deviation from the BPS limit.  Using
the general Kaluza-Klein ansatz to lift to a solution of 11d
SUGRA~\cite{manyauthors}, the charged black hole solution (\ref{7dbh})
becomes
\begin{eqnarray*}
ds_{11}^2& = & \Delta^{1/3}\Bigl((H_1 \, H_2)^{-1}f\,
dt^2+(f^{-1}dr^2+r^2 d\Omega_5^2)\Bigr)\\ && \,\,\,\,\,\,\,
+\frac{1}{4} \Delta^{-2/3}\Bigl[ d\mu_0^2+\sum_{i=1}^2
H_i\Bigl(L_7^2 \, d\mu_i^2+\mu_i^2(L_7 \, d\phi_i+
\frac{\tilde{q}_i}{r^4+q_i}dt)^2\Bigr)\Bigr] \, ,
\end{eqnarray*}
where
\begin{equation}
   \Delta \equiv (H_1 H_2)\Bigl(\mu_0^2+\sum_{i=1}^2 \frac{\mu_i^2}{H_i}\Bigr),  ~~~~
   \mu_0 \equiv \sin\theta_1 \sin \theta_2, ~~~~ \mu_{1} \equiv \cos\theta_{1},~~~~ \mu_{2} \equiv \sin\theta_{1}\,
   \cos\theta_{2}\,,
\end{equation}
and $L_7^2=4 l_p^2 (\pi N)^{2/3}$.  This solution is
asymptotically $\ads{7} \times \sph{4}$ with scales $L_7$ and
$L_7/2$ respectively. $N$ counts the number of units of background
7-form flux, or equivalently, the number of M5-branes whose
near-horizon limit is $\ads{7}\times \sph{4}$ with $\ads{7}$ scale
$L_7$. First consider the BPS solution ($\mu=0$).  Following the
reasoning in~\cite{myers} and the previous section, the relevant
7-form component is
\begin{equation}
F^{(7)}_{\rho_i \phi_i \alpha_1 \alpha_2 \alpha_3 \alpha_4
\alpha_5 }= 4 \, {q}_i \, \rho_i \,
\sin^4\alpha_1\,\sin^3\alpha_2\,\sin^2\alpha_3\,\sin\alpha_4\,,
\end{equation}
where $\rho_i\equiv {L_7 \over 2} \sqrt{1-\mu_i^2}$. The density of
giant gravitons is
\begin{equation}
\frac{dn_i}{d\rho_i}=\frac{1}{16\pi G_{11} T_2}
   \int F^{(7)}_{\rho_i \phi_i \alpha_1
\alpha_2 \alpha_3 \alpha_4 \alpha_5 } d\phi_i d^5 \alpha =8
N^{2}\frac{q_i}{L_7^6} \rho_i \,. \label{7dn}
\end{equation}
The total number of giant gravitons in each set is
\begin{equation}
n_i=\int_0^{L/2} d\rho_i 8 N^{2}\frac{q_i}{L_7^6} \rho_i=
N^{2}\frac{q_i}{L_7^4} \,. \label{totnum7}
\end{equation}
Giant gravitons of radius $\rho_i$ carry angular momentum
$\frac{2N}{L_7} \rho_i$~\cite{giant} and therefore the total angular
momentum carried by each set of giant gravitons is
\begin{equation}
P_{\phi_i}=\int_0^{L/2}d\rho_i \, \frac{dn_1}{d\rho_i}
\frac{N}{L_7} \rho_i=\frac{2N^{3}}{3}\frac{q_i}{L_7^4}\,.
\label{7dmom}
\end{equation}
Likewise, using the energy-momentum relation for giants, the total
energy of the condensate of giants is
\begin{equation}
E=\frac{P_{\phi_1}+P_{\phi_2}}{L_7/2}=\frac{4N^{3}}{3}\frac{q_1+q_2}{L_7^5}
\, . \label{E7}
\end{equation}
As before, these energy and momentum calculations are performed while
treating giant
gravitons as probes in a background $\ads{7} \times \sph{4}$
geometry.   Nevertheless, the total energy agrees exactly with the
spacetime mass of the solution~(\ref{7dmass})
\begin{equation}
M_7=\frac{\pi^2}{4G_7}(q_1+q_2)=\frac{4N^{3}}{3}\frac{q_1+q_2}{L_7^5}
\, .
\end{equation}
where we have used $G_7=6G_{11}/(\pi^2 L_7^4)$.
In contrast, the non-supersymmetric solutions with $\mu>0$ have a net
condensate of giants in which $q_{i}$ in (\ref{7dn}) and
(\ref{7dmom}) is replaced by the physical charge $\tilde{q}_{i}$ defined as in
(\ref{physch4}).   A similar replacement in the giant energy
formula (\ref{E7}) will not reproduce the mass (\ref{7dmass}) because
at positive $\mu$ additional excitations of the giants as well as
brane-anti-brane pairs may be present.

\section{Entropy}

Above we have generalized the calculation of Myers and
Tafjord~\cite{myers} in 5d to interpret certain charged black
holes and curvature singularities of 4d and 7d gauged supergravity
as condensates of giant gravitons.  Giant gravitons are themselves
spherical M2, D3 and M5 branes, and so we expect that after
including gravitational back-reaction the geometry very close to
the surface of a giant should be locally $\ads{l} \times \sph{k}$.
In particular, near the M5-brane giants of the $\ads{4} \times
\sph{7}$ space, the geometry should be locally $\ads{7} \times
\sph{4}$.  Likewise, near the M2-brane giants of $\ads{7} \times
\sph{4}$ the local geometry should be $\ads{4} \times \sph{7}$.
Finally, near the D3-brane giants of $\ads{5} \times \sph{5}$ the
local geometry should again be $\ads{5} \times \sph{5}$, but the
AdS scale should be determined by $N_g$, the number of giants,
rather than $N$, the number of background branes.  This suggests a
simple correspondence principle that should hold when a single
species of giant is condensed: {\it When the number of giants in
$\ads{k} \times \sph{l}$ is sufficiently large, the physics can be
equivalently described by a dual $\ads{l} \times \sph{k}$ in which
the giants and the background branes have exchanged roles}.

If this correspondence is correct, the thermodynamics of the large
charge black holes of $\ads{k} \times \sph{l}$ and the small
charge black holes of $\ads{l} \times \sph{k}$ should be
equivalent.  In the near-BPS limit we expect it to imply equality
of black hole entropies and temperatures.  Below, we test this by
expressing the entropy and temperature of near-BPS, single charge
black holes of $\ads{k} \times \sph{l}$ in terms of the number of
condensed giants $N_g$, the number of background branes $N$ and
the energy above extremality $\delta M$.


\paragraph{\boldmath$\ads{5} \times \sph{5}$: } The previous sections have
discussed the charged black hole solutions of $\ads{7}$ and
$\ads{4}$ but here we begin by analyzing the entropy of the
$\ads{5}$ black holes whose interpretation in terms of giant
gravitons was given in~\cite{myers}.  In detail, the $\CN = 8, \,
SO(6)$ gauged SUGRA arising from consistent truncation of 10d, IIB
SUGRA compactified on $\sph{5}$ has a further truncation to 5d
$\CN=2, \, U(1)^{3}$ gauged SUGRA. The black hole solutions of
this theory are ~\cite{5dholes}
\begin{eqnarray}
ds_5^2&=&-(H_1H_2H_3)^{-2/3} f \, dt^2 + (H_1H_2H_3)^{1/3}
(f^{-1}dr^2+r^2 d\Omega_{3}^2), \nonumber \\ X_i& = &
H_i^{-1}(H_1H_2H_3)^{1/3}, \,\,\,\,\,\,\,\,
A^i=\frac{\tilde{q}_i}{r+q_i} \, dt,  ~~~~~(i = 1\cdots 4)
\label{5dbh}
\end{eqnarray}
where
\begin{equation}
f=1-\frac{\mu}{r^2}+\frac{r^2}{L_5^2}(H_1H_2H_3), ~~~~~~
H_i=1+\frac{q_i}{r^2}.
\end{equation}
The solution has a mass
\begin{equation}
M_5 = {\pi \over 4G_{5}}( {3 \over 2} \pi \mu + \sum q_{i}) = {\pi
\over 4G_{5}}\sum q_{i} + 3\, {\delta} M_5 \label{deltam}\,.
\end{equation}
Here $3\delta M_5$ measures the additional mass over the BPS limit
produced by turning on $\mu$.  In the single charge BPS case ($\mu
= 0$) Myers and Tafjord showed that the singularity contains
\begin{equation}
N_{g} = N \, {q  \over L_5^{2}} \label{ng5d}
\end{equation}
giants where $ L_5^4=4\pi g_s N l_s^4 $ and $N$ is the number of
background units of 5-form flux, or equivalently the number of
3-branes whose near-horizon limit gives $\ads{5}$ with scale size
$L_5$.  Beyond the BPS limit ($\mu > 0$), the number of giants is
given by replacing $q$ in (\ref{ng5d}) by the physical charge
$\tilde{q}$ as in (\ref{physch4}). The horizon of the single
charge black holes occurs at
\begin{equation}
r_h=\frac{1}{2}\sqrt{-2(L_5^2+q)+2\sqrt{(L_5^2+q)^2+4\mu L_5^2}} \,.
\end{equation}
We will consider near-BPS limits in which $\mu \ll q  $ and
compute the horizon entropy in the two cases $q\ll L_5^{2}$ and
$q\gg L_5^{2}$, or equivalently $N_{g} \ll N$ and $N_{g} \gg N$.
Gravitational entropy of a 5d horizon is given by
\begin{equation}
S=\frac{A}{4G_5}=\frac{\Omega_3}{4G_5}r_h^2\sqrt{r_h^2+q} \,,
\end{equation}
where $r_h$ is the location of the horizon,
$G_5=\frac{G_{10}}{\pi^3L_5^5}$ and $G_{10}=8\pi^6g_s^2l_s^8$ are
the 5d and 10d Newton constants.\footnote{$\Omega_{D} =
{2\pi^{(d+1)/2} \over \Gamma[(d+1)/2]} $ is the volume of the unit
D-sphere.} In terms of the number of giants we find that
\begin{eqnarray}
\begin{array}{|c| c| c|}
\hline {\rm Limits} &   r_h  &  S=\frac{A}{4G_4}  \\ \hline \mu\ll
q,~ N_{g}\ll N &     \sqrt{\mu}  & S = {4\pi} (L_5 \, \delta M_5 )
\sqrt{{N_{g}
    \over N}}
  \\ \hline
\mu\ll q,~N_{g}\gg N
   &     r_{h} =  L_5 \, \sqrt{\mu / q}
 &
S = {4 \pi} (L_5 \, \delta M_5 )
       \sqrt{{N \over N_{g}}}
\\\hline
\end{array}
\label{5dentropies}
\end{eqnarray}
Notice that in the small and large charge limits, the entropy
formula interchanges the roles of the giant gravitons ($N_{g}$)
and the D3-branes ($N$) whose near-horizon created the background
geometry, if we hold $L_5$ and $\delta M_5$ fixed.

\paragraph{\boldmath$\ads{4} \times \sph{7}$: }
We will study single charge 4d black holes (\ref{4dbh}) for which
the horizon occurs when
\begin{equation}
f = 1-\frac{\mu}{r}+\frac{qr}{L_4^2}+\frac{r^2}{L_4^2}=0\,.
\end{equation}
As above we will compare the entropy of small charge ($q\ll L_4$)
and large charge ($q \gg L_4$) black holes.  From (\ref{4dtot1}),
the number of giants $N_{g}$ satisfies $N_{g} \ll \sqrt{N}$ and
$N_{g} \gg \sqrt{N}$ in these cases.  We will study the near-BPS
solutions for which $\mu \ll q$, but in the large charge case will
need to separately consider the regimes $\mu \ll L_4^{2}/q$ (${\mu
\over L_4} \ll {\sqrt{N} \over N_g}$) and $L_4^{2}/q \ll \mu \ll
q$ (${\sqrt{N}\over N_g} \ll {\mu \over L_4}\ll {N_g\over
\sqrt{N}}$).  The gravitational entropy is
\begin{equation}
S=\frac{A}{4G_4}=\frac{\Omega_2}{4G_4} \, r_h^2 \,
\sqrt{1+\frac{q}{r_h}}\,,
\end{equation}
where $r_h$ is the location of the horizon. Then, using $G_{4} =
G_{11}/{\rm Vol}(\sph{7})$ and $G_{11}$  as in Sec.~2,  we obtain,
\begin{eqnarray}
\begin{array}{|c| c| c|}
\hline {\rm Limits} &   r_h  &  S=\frac{A}{4G_4}  \\ \hline
\frac{\mu}{L_4} \ll \frac{N_g}{\sqrt{N}},~ N_{g}\ll \sqrt{N} &
\mu  &
 \sqrt{48 \pi^{2}} \, (L_4 \, \delta M_4 )^{3/2} \, {\sqrt{N_g} \over N}   \\ \hline
     {\mu \over L_4} \ll {\sqrt{N} \over N_g},~ N_{g}\gg \sqrt{N} &     \mu  &
 \sqrt{48 \pi^{2}} \, (L_4 \, \delta M_4 )^{3/2} \, {\sqrt{N_g} \over N}
\\\hline
  {\sqrt{N}\over N_g} \ll {\mu \over L_4}\ll {N_g\over \sqrt{N}},~N_{g}\gg \sqrt{N} &   L_4 \,
    \sqrt{\mu/q} &
   4\pi
     {1 \over 3^{1/4}} \, (L_4 \, \delta M_4)^{3/4} \,
     {\sqrt{N} \over N_{g}^{1/4}} \\ \hline
\end{array}
\label{4dentropies}
\end{eqnarray}
Here $\delta M_4$ is a measure of energy over the BPS limit as in
(\ref{MADM4}).  Unlike the $\ads{5}$ case the entropies for the
small and large charge cases do not seem to be related to each
other. However, we will see a remarkable fact below --  the large
charge $\ads{4}$ black hole entropy is related to the small charge
$\ads{7}$ entropy and vice versa.

\paragraph{\boldmath $\ads{7} \times \sph{4}$: }
The horizon of the single charge 7d black hole (\ref{7dbh}) occurs
when
\begin{equation}
1-\frac{\mu}{r^4}+\frac{r^2}{L_7^2}+\frac{q}{r^2L_7^2}=0\,.
\end{equation}
We will consider  small charge ($q \ll L_7^{4}$) and large charge
($q \gg L_7^{4}$) limits, which imply  $N_{g} \ll N^{2}$ and
$N_{g} \gg N^{2}$ respectively from(\ref{totnum7}). As above, we
will study the near-BPS ($\mu \ll q$) limit, but in the small
charge case we must separately consider the cases
$\Bigl(q/L_7^{2})^{2} \ll \mu \ll q$ $~\Bigl({N_g^2 \over N^4} \ll
{\mu \over L_7^4} \ll {N_g \over N^2}\Bigr)~$ and $~\mu \ll
(q/L_7^{2})^{2}$ $~\Bigl({\mu \over L_7^4} \ll {N_g^2 \over
N^4}\Bigr)$. The gravitational entropy is given by
\begin{equation}
S=\frac{A}{4G_7}=\frac{\Omega_5}{4G_7}\sqrt{1+\frac{q}{r_h^4}}r_h^5 \,.
\end{equation}
Using $G_{7} = G_{11}/{\rm Vol}(\sph{4})$ and $G_{11}$ as in
Sec.~3,
\begin{eqnarray}
\begin{array}{|c|c|c|}
\hline {\rm Limits} & r_h& S=\frac{A}{4G_7} \\ \hline
    {\mu \over L_7^4} \ll {N_g \over N^2} ,~N_{g}\gg N^{2}&   L_7 \, \sqrt{\mu/q}&
 \sqrt{48\pi^2 }\,
(L_7 \, \delta M_7)^{3/2} \, \frac{\sqrt{N}}{N_{g}} \\ \hline
    {\mu \over L_7^4} \ll {N_g^2 \over N^4} ,~ N_{g}\ll N^{2}&  L_7 \,
    \sqrt{\mu/q}&
  \sqrt{48\pi^2} \,
(L_7 \, \delta M_7)^{3/2} \, \frac{\sqrt{N}}{N_{g}} \\ \hline
  {N_g^2 \over N^4}  \ll {\mu \over L_7^4} \ll {N_g \over N^2},~ N_{g}\ll N^{2}&  \mu^{1/4} &
    4 \pi {1 \over  3^{1/4}} \, (L_7 \, \delta
    M_7)^{3/4}\, {\sqrt{N_{g}} \over N^{1/4}} \\ \hline
\end{array}
\label{7dentropies}
\end{eqnarray}
Here $\delta M_7$ is a measure of energy over the BPS limit as in
(\ref{7dmass}).

\subsection{A thermodynamic correspondence}

The entropy formulae derived above were necessarily functions of
$\delta M$, the energy above extremality.  However, since entropy
is itself dimensionless, this energy always appears in the
dimensionless combination $L \, \delta M$, where $L$ is the AdS
scale.  The appearance of this combination, which is natural from
the point of the gravitational solution, also has a natural
interpretation from the microscopic point of view.  Das, Jevicki
and Mathur have shown that a giant graviton has a discrete
spectrum that is determined by the AdS scale $L$ and is {\it
independent} of all other length scales including the radius of
the giant~\cite{fluct}.  This remarkable fact, coupled with the
appearance of $L \, \delta M$ in the above entropy formulae,
suggests that the thermodynamics of the charged black holes
studied here can be microscopically explained in terms of the
fluctuations of giant gravitons that make up the singularity.  In
effect, the giant gravitons appear to play a role for our black
holes analogous to the role of the D1-D5 string in the classic
Strominger-Vafa analysis of black hole entropy in string
theory~\cite{stromvaf}.

Above we argued that when the number of giants in the solution is
large there ought to be a correspondence exchanging the roles of
the giant gravitons and the background branes creating the AdS
spacetime. Essentially, this should relate the entropies of large
charge black holes in $\ads{l}\times \sph{k}$ and small charge
black holes in $\ads{k} \times \sph{l}$.  The results presented in
(\ref{5dentropies}), (\ref{4dentropies}) and (\ref{7dentropies})
are a striking confirmation of this fact.  In $\ads{5}$ the
entropy at large $N_g$ is obtained from the small $N_g$ result by
exchanging $N$ and $N_g$ while holding the dimensionless energy
$L_5 \, \delta M_5$ fixed.  Similarly, the large $N_g$ entropy in
$\ads{4}$ reproduces the small $N_g$ entropy in $\ads{7}$ if we
hold the dimensionless energy fixed ($L_4 \, \delta M_4 = L_7 \,
\delta M_7$) while exchanging $N_g$ and $N$.  The numerical
factors match exactly, as do the limits on the supersymmetry
breaking parameter $\mu$.

We can derive the temperature of our solutions from the
thermodynamic relation $\beta = 1/T = dS/dE$.  By requiring that
the temperatures match between corresponding solutions in $\ads{l}
\times \sph{k}$ and $\ads{k} \times \sph{l}$ we find that while
exchanging $N$ and $N_g$ we should separately hold $\delta M$
and the supergravity length scale $L$ constant.  Since $L$ can be
expressed in terms of the number of background branes and the
fundamental length scales $l_5 = g_s^{1/4} l_s$ or the Planck
length $l_p$, this requires us to rescale $l_5$ or $l_p$ when
$N_g$ and $N$ are exchanged.  In particular, $N_g \gg N$ solutions
in $\ads{5}$ with a given value of $l_5 = g_s^{1/4} l_s$ are
related to the $N_g \ll N$ solutions with $\tilde{l}_5 = l_5
(N/N_g) \ll l_5$.  Similarly, the $N_g \gg \sqrt{N}$ solutions in
$\ads{4}$ with a given Planck length $l_p$ are related to $N_g \ll
N^2$ spaces in $\ads{7}$ with $\tilde{l}_p = l_p (1/2)^{7/6}
(N/N_g^2)^{1/6} \ll l_p$.  Finally, the $N_g \gg N^2$ solutions in
$\ads{7}$ with a given $l_p$ are related to $N_g \ll \sqrt{N}$
black holes in $\ads{4}$ with $\tilde{l}_p = l_p (2)^{7/6}
(N^2/N_g)^{1/6} \ll l_p$.  In all cases, the new fundamental
length scale is much smaller than the original scale.  While this
will be important for a quantum mechanical analysis of our
proposed correspondence, it is not relevant for the supergravity
analysis of this article since only the scale $L$ appears in the
spacetime solutions.

\section{Discussion: A correspondence principle?}

We have presented evidence from supergravity that there is a
correspondence between the physics of certain large charge black
holes in $\ads{l}\times\sph{k}$ and small charge black holes in
$\ads{k}\times\sph{l}$.  It would be interesting to extend the
correspondence that we suggest to a full quantum mechanical
duality.

To this end, it is important to understand how the charged black
holes discussed here are represented in the CFT dual to AdS space.
In~\cite{gkpw,bkl} it was shown that non-normalizable bulk modes
(boundary conditions) map to CFT couplings, while normalizable
supergravity lumps give rise to VEVs for CFT operators.  The
operators relevant to condensation of giant gravitons were
identified in~\cite{subdet} as subdeterminants of products of CFT
fields. Presumably, the black holes we have studied are related to
superselection sectors of the CFT in which these operators have VEVs.

The CFT state corresponding to large $N_g$ would not be
well-approximated by the planar limit. (See~\cite{subdet} for a
discussion of the combinatorial explosion in perturbative
calculations involving giant gravitons.)  In the bulk spacetime
non-planar diagrams correspond to string loop corrections.  The
correspondence we are proposing could be understood in $\ads{5}$
as stating that these loop corrections re-sum to give another
description in which the $SU(N)$ gauge symmetry of
the dual CFT is exchanged with an $SU(N_g)$ symmetry arising from
the condensed giants.

A brief glance at the entropy formulae in (\ref{5dentropies}),
(\ref{4dentropies}) and (\ref{7dentropies}) and the relation $1/T
= dS/dE$ shows that the black holes that we are examining have
unusual thermodynamic relations.  For example, in the limits we
examined in (\ref{5dentropies}) the temperature is independent of
the energy.  In fact, some of the solutions we have considered
here have thermodynamic, and possibly dynamical,
instabilities~\cite{therm,instabilities}.  We expect that such
solutions with large charges in $\ads{k} \times \sph{l}$ will
correspond to unstable small charge solutions in $\ads{l} \times
\sph{k}$ and that the instabilities can be related to each other.
Understanding this will be important for the goal of relating
various features of the gravitational solutions to the physics of
the giant gravitons making up the singularities.

\vspace{0.25in} {\leftline {\bf Acknowledgements}}

We thank Micha Berkooz and Matt Strassler for useful discussions.
This work was supported by DOE grant DOE-FG02-95ER40893.

\end{document}